# Hierarchical automation of scanning probe microscopy through agentic orchestration and algorithmic control

Boris N. Slautin,[1,*,†] Sheryl L. Sanchez,[1,*,†] Aidan Swanger,[1] Yu Liu,[1]
Gerd Duscher,[1] Vladimir V. Shvartsman,[2] Mahshid Ahmadi,[1]
Sergei V. Kalinin[1,*]

[1] Department of Materials Science and Engineering, University of Tennessee; Knoxville, Tennessee, 37996, USA
[2] Institute for Materials Science and Center for Nanointegration Duisburg-Essen (CENIDE), University of Duisburg-Essen, Essen, 45141, Germany

## Abstract

Rapid advances in agentic artificial intelligence enable scientific systems to interpret open-ended objectives, combine heterogeneous information, invoke specialized tools, and revise experimental strategies as evidence accumulates. However, physical experimentation also contains many tasks for which agentic reasoning provides little advantage and can reduce reliability. Quantitative analysis, optimization, spatial targeting, validation, and instrument execution are often better posed as deterministic or algorithmic operations with explicit objectives and verifiable outputs. Here, we introduce a hierarchical architecture for autonomous experimentation that separates these roles. Agentic components interpret scientific intent, construct task-dependent experimental representations, evaluate accumulated evidence, and select high-level actions, whereas deterministic algorithms perform numerical analysis, coordinate selection, validation, and physical execution. We implement this architecture in piezoresponse force microscopy. Starting from a broad scientific question concerning the relation between local domain structure and polarization switching, the system constructs spatial descriptors from multichannel imaging, selects and analyzes local hysteresis measurements, adapts the spectroscopy waveform, and terminates the experiment when additional measurements cease to provide new evidence. The autonomous trajectory also identifies a confounding relationship between polarization state and domain-wall proximity and recognizes that the requested contrast is not independently represented within the available field of view. These results demonstrate a route toward scientific autonomy in which agents determine what evidence is required while algorithms determine how that evidence is acquired reproducibly and within validated physical constraints.

[*] Authors to whom correspondence should be addressed: bslauti1@utk.edu, ssanch18@vols.utk.edu, and sergei2@utk.edu
[†] These authors contributed equally to this work.

## 1. Introduction

Experimental automation has progressed from scripted measurement sequences to closed-loop systems in which acquired data determine subsequent actions.[1-6] In chemistry and materials science, autonomous laboratories have combined robotic synthesis and characterization with active learning and Bayesian optimization (BO) for phase mapping, materials discovery, and process or formulation optimization.[7-11] Gaussian-process and BO methods have also been developed and benchmarked for experimental selection, quality control, and human oversight.[7, 12, 13] In microscopy, related approaches have enabled autonomous measurement selection, structure–property discovery, combinatorial-library exploration, and instrument optimization.[14-18] In many such implementations, the operative objective or acquisition function and the available action vocabulary are externally specified. These methods are therefore particularly effective when the scientific problem can be represented through a defined descriptor space, objective, and set of admissible actions.

Agentic artificial intelligence adds a different capability. Scientific agents can interpret natural-language objectives, combine heterogeneous information, invoke computational and experimental tools, and modify multistep workflows as new information becomes available.[19-39] This creates the possibility of experiments in which scientific intent is propagated through planning, analysis, measurement, and control rather than encoded completely in advance.

Purely agentic control, however, is neither necessary nor sufficient for physical experimentation. Language-model reasoning can be brittle for numerical and low-level spatial tasks, can vary between nominally equivalent runs or prompt formulations, and can produce unsupported statements.[40-42] Tool use mitigates these limitations by delegating calculations and specialized operations to external functions, programs, and cooperating components, while autonomous scientific systems also require explicit safety boundaries and task-level evaluation.[21, 23, 43-45] More fundamentally, many experimental operations already have a well-defined mathematical objective, including fitting a response, optimizing a known acquisition function, selecting the extremum of a spatial criterion, validating an instrument state, or enforcing hardware limits. For these operations, an agent often adds only a superficial reasoning layer around a task that a dedicated algorithm can perform more reproducibly and efficiently.

The central question is therefore not whether an autonomous experiment should be agentic or algorithmic, but which representation and control mechanism should be assigned to each task and how the interaction between top-down agentic and bottom-up optimization workflow should be enabled.[46] Here, we develop a hierarchical dual agentic-algorithmic architecture in which agents are used for decisions that depend on scientific intent and evolving experimental context, whereas numerical analysis, optimization, coordinate selection, validation, and physical execution remain deterministic or algorithmic. We then implement this architecture in scanning probe microscopy (SPM) and demonstrate it in a closed-loop piezoresponse force microscopy (PFM) experiment.

## 2. Hierarchical agentic-algorithmic architecture for autonomous experiments

### 2.1. Task allocation between agents and algorithms

Physical experiments involve distinct operations that differ in how information is represented, how measurement parameters are selected, and how the instrument is controlled. Examples include numerical filtering, descriptor extraction, model fitting, constrained optimization, coordinate selection, validation against hardware limits, and execution of instrument commands. Once the relevant quantity or objective has been specified, these operations are naturally handled by deterministic algorithms or specialized optimizers.

Other operations are conditional on the scientific objective and on the interpretation of accumulated evidence to construct fixed-policy or reward driven workflows. The same image may require different representations depending on whether the experiment concerns domain walls, grain boundaries, local contact mechanics, or acquisition artifacts. Likewise, whether another measurement is useful can depend on which scientific distinctions remain unresolved, whether experimental variables are confounded, and whether the accessible state space contains the contrast required to answer the question.

We therefore assign tasks according to whether the required input-output relation can be specified and verified independently of the evolving scientific context. Agentic nodes determine what information is scientifically relevant and what evidence should be acquired next. Algorithmic nodes determine how defined quantities are calculated, optimized, validated, and physically executed. This distinction is not equivalent to a separation by experimental timescale: a high-level optimization problem may be entirely algorithmic if its objective is explicit, whereas interpretation of a single measurement may require contextual reasoning if its significance depends on the scientific question.

### 2.2. Experimental hierarchy and shared evidence state

The task-allocation principle is general and can be implemented as a recursive experimental graph, illustrated here for PFM (Figure 1). A scientific task and experimental context initialize the experiment. Measurements are acquired through validated instrument interfaces and routed to measurement-specific analysis node. This node converts raw observations into compact structured evidence. A decision node evaluates this evidence together with the experimental history and determines the next high-level action. The cycle continues until the scientific task is addressed or a predefined resource or safety limit is reached.

**1. Goal, task, and context definition.** The system receives the scientific question, available prior information, instrumental constraints, and experimental budget.

**2. Evidence acquisition.** A requested experimental action is translated into validated instrument operations. Physical execution is deterministic and bounded by predefined hardware limits.

**3. Reward and state construction.** Measurement-specific analysis converts acquired data into representations appropriate for subsequent scientific decisions. Quantitative transformations remain algorithmic, whereas task-dependent interpretation can invoke an agent.

**4. Decision.** The accumulated evidence is evaluated with respect to the scientific objective. The system determines what remains unresolved and selects the next type of measurement or terminates the experiment.

A shared *Experimental state* links these operations. It contains the scientific task and context, instrument status, measurement history, generated representations, quantitative analysis results, previous decisions, identified limitations, and remaining budget. High-dimensional numerical arrays remain outside the orchestration state and are accessed through validated references. Consequently, high-level decisions operate on structured evidence rather than unrestricted raw microscope data.

This architecture also defines the physical-control boundary. An agent can request a scientifically meaningful action, such as measuring a wall-proximal region or increasing a bias range after an unsaturated loop, but it does not directly calculate the final tip coordinate or issue unbounded hardware commands. Those operations are delegated to validated algorithms and instrument interfaces.

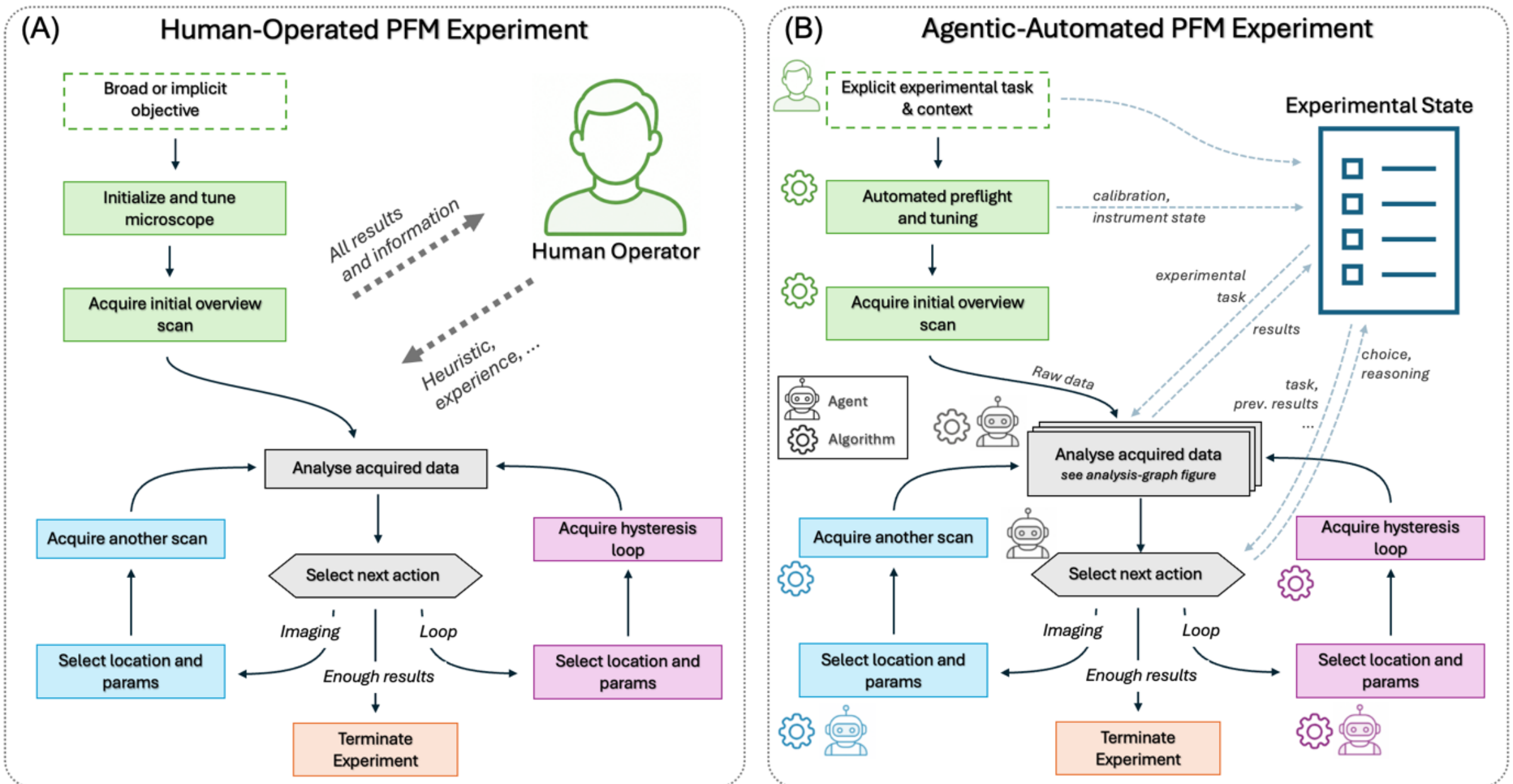

**Figure 1. Hierarchical agentic-algorithmic experimental workflow.** (A) A human-operated PFM experiment uses prior knowledge to interpret each measurement and select subsequent imaging, local spectroscopy, or termination. (B) In the automated workflow, an explicit task and context guide agentic interpretation and high-level decision-making, while deterministic nodes execute validated analysis, coordinate selection, and instrument operations. A shared experimental state records accumulated evidence and closes the acquisition-analysis-decision loop.

The analysis node is itself organized as a nested graph with separate pathways for multichannel imaging and local hysteresis spectroscopy (Figure 2). The image branch converts the acquired multichannel images into task-dependent spatial criteria to select feature of interest for subsequent spectroscopy. Local hysteresis loops are first reduced deterministically to quantitative descriptors and subsequently evaluated by an agent with respect to measurement quality and physical interpretation. Both branches return structured evidence to the shared Experimental state.

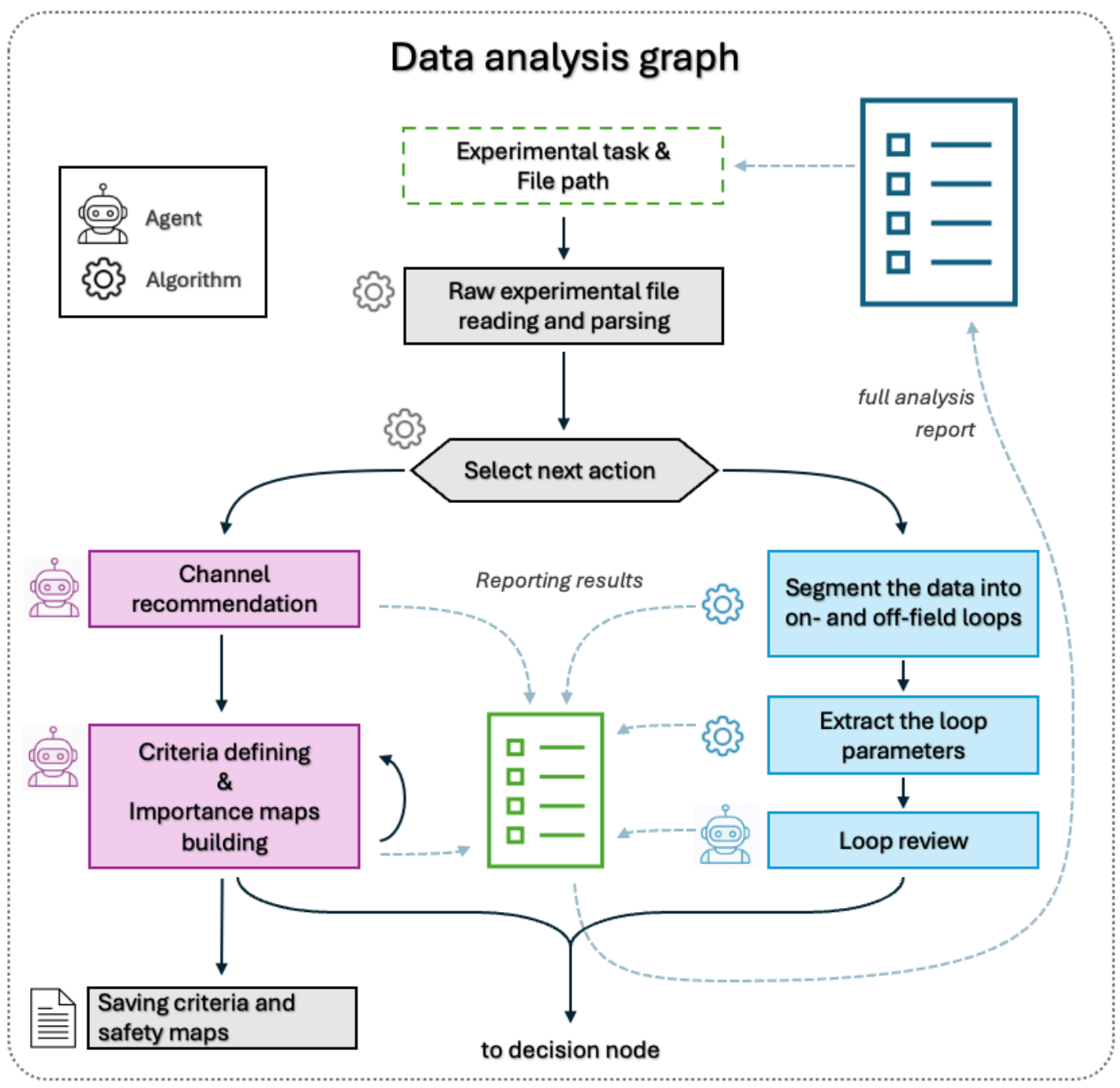

**Figure 2. Agentic-algorithmic data-analysis graph for imaging and local hysteresis spectroscopy.** Every acquired measurement is read and classified deterministically, then routed to an image-analysis or loop-analysis subgraph. Image analysis constructs task-dependent spatial criteria and an acquisition-validity map; loop analysis extracts quantitative descriptors before agentic quality assessment and physical interpretation. Both branches return structured evidence to the experimental state.

## 2.2. Task-conditioned representations as the interface between reasoning and optimization

A central problem in autonomous experimentation is translating a broad scientific question into quantities that can actually be measured or optimized. Raw experimental channels rarely correspond one-to-one with the concepts appearing in a scientific question. Instead, the relevant representation often must be constructed from several observations.

Here we use agentic reasoning to construct task-conditioned experimental coordinates dynamically during the experiment. Given the scientific objective and the available measurements, the agent identifies the physical distinctions relevant to the task and defines computational procedures that map the acquired data onto corresponding spatial or scalar descriptors. Once these quantities are defined, their numerical evaluation and optimization remain algorithmic. The resulting hierarchy connects the scientific objective to a task-conditioned representation, which guides algorithmic optimization and physical measurement, producing quantitative evidence that supports scientific interpretation.

# 3. Implementation in autonomous PFM

Modern microscopy techniques increasingly combine high-dimensional imaging, spectroscopy, and spatially resolved measurements, requiring autonomous systems to integrate data interpretation, quantitative analysis, measurement selection, and instrument control. As a model system, we chose PFM that combines spatially resolved imaging with in-point spectroscopy and provides a testbed for developing and evaluating autonomous experimental workflows.[47, 48] Within this technique, DART-PFM provides resonance-tracked electromechanical contrast,[49] while switching-spectroscopy PFM (SS-PFM) provides hysteresis responses from which coercive bias, imprint, remanent response, and related switching descriptors can be extracted quantitatively.[50]

The autonomous PFM experiment implemented here begins with an overview DART-PFM scan, from which the workflow constructs task-relevant spatial criteria and selects locations for local switching spectroscopy. Each acquired loop is analyzed and added to the experimental state, after which the system determines whether additional spectroscopy, a new scan, modification of the spectroscopy protocol, or termination is required. In the end-to-end demonstration, the workflow is initialized with the broad task of determining how local domain structure affects polarization switching dynamics, without predefined spatial criteria, measurement locations, spectroscopy waveform, or stopping rule.

### 3.1. PFM imaging from a scientific question to spatial experimental coordinates

Multichannel PFM imaging contains heterogeneous information about surface morphology, electromechanical response, polarization-related contrast, and local tip-sample contact conditions. In the present workflow, imaging is not the final measurement step. Its role is to define spatial variables that make the scientific question experimentally actionable, meaning select the locations for the spectroscopic studies based on the observed image features. Note that these decisions cannot be made prior to image observation.

The channel-recommendation node evaluates which acquired channels contain information relevant to the requested analysis without itself performing segmentation (Figure 3). In the lead lanthanum zirconate titanate (PLZT) ceramic PFM scan (Figure 3B), phase and amplitude were identified as the primary sources of domain-related information, height as the relevant morphological channel, and resonance-frequency contrast primarily as an indicator of contact or acquisition conditions. Repeated evaluations returned stable primary-channel assignments, while confidence decreased for tasks such as grain-boundary identification for which the available channels contained little explicit contrast (Figure 3C,D).

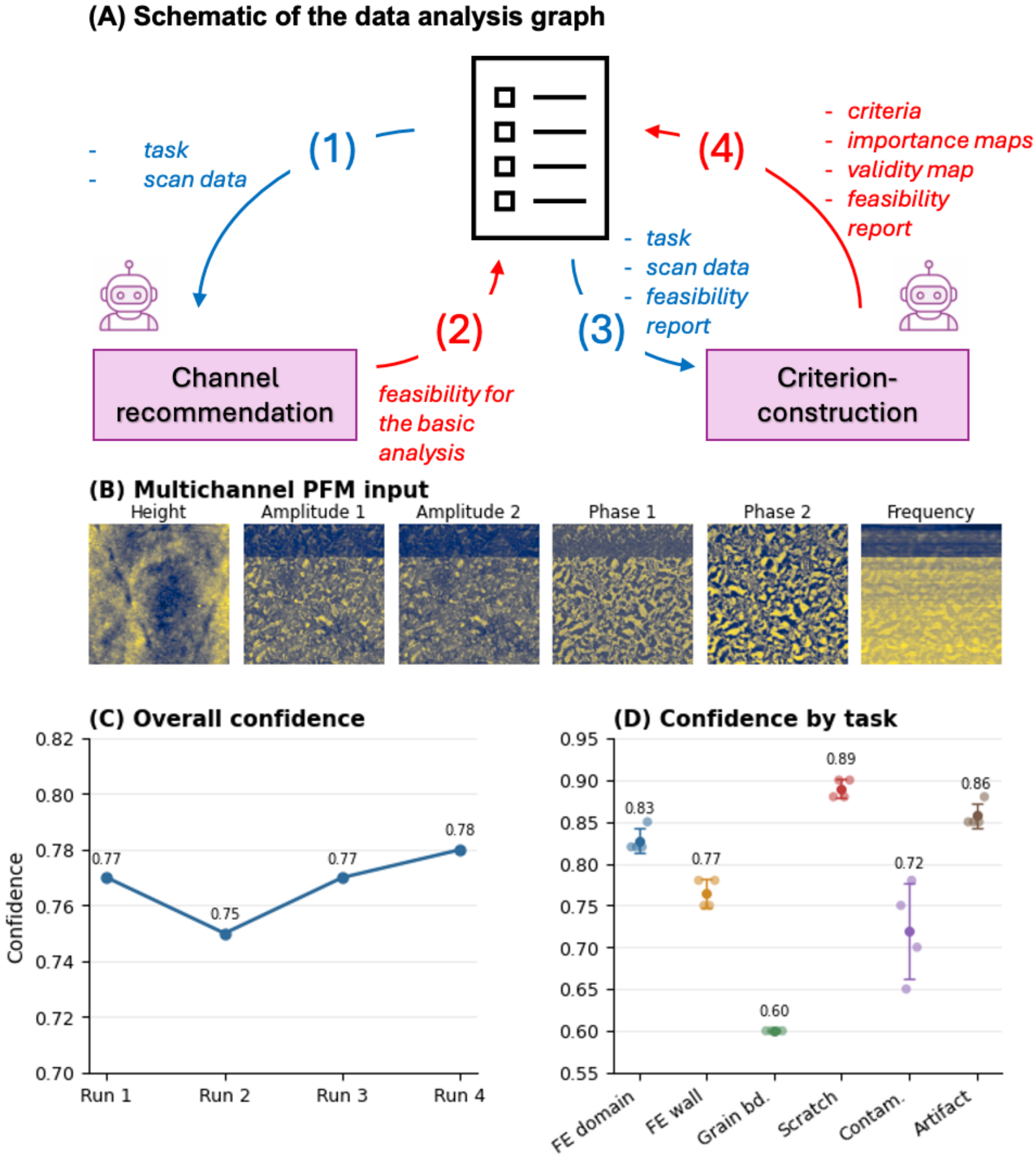

**Figure 3. Agentic channel assessment for multichannel PFM analysis.** (A) Image-analysis workflow with channel recommendation and criterion construction. (B) Multichannel DART-PFM dataset. (C) Repeated evaluation of overall confidence. (D) Task-dependent confidence, with lower feasibility for analyses weakly supported by the acquired contrast.

Given the scientific task, the criterion-construction node converts the relevant image channels into task-conditioned spatial descriptors (Figure 4A). Across repeated analyses, the dominant semantic categories were domain state or response, domain-wall proximity, and contact- or stiffness-related variation; local domain complexity or structural-boundary descriptors appeared less consistently (Figure 4B). The image branch independently generated an acquisition-validity map identifying regions where spectroscopy could be compromised by large topographic gradients, surface particles, contact instabilities, or scan artifacts (Figure 4D). This map is treated separately from the scientific criteria and constrains subsequent coordinate selection by reducing or excluding locations where reliable spectroscopy is unlikely.

The scientific concepts were more reproducible than their exact pixel-wise numerical realizations (Figure 4B, C). Different runs frequently generated different computational implementations of the same physical concept, whereas the same dominant physical variables recurred. This distinction is important as the agent is constructing both a scientific quantity and an executable approximation to it from the available channels. Reproducibility should therefore be evaluated at the level of experimentally consequential distinctions as well as at the level of pixel-wise agreement. Additional analysis of the agentic PFM scan processing is provided in the Supplementary Information.

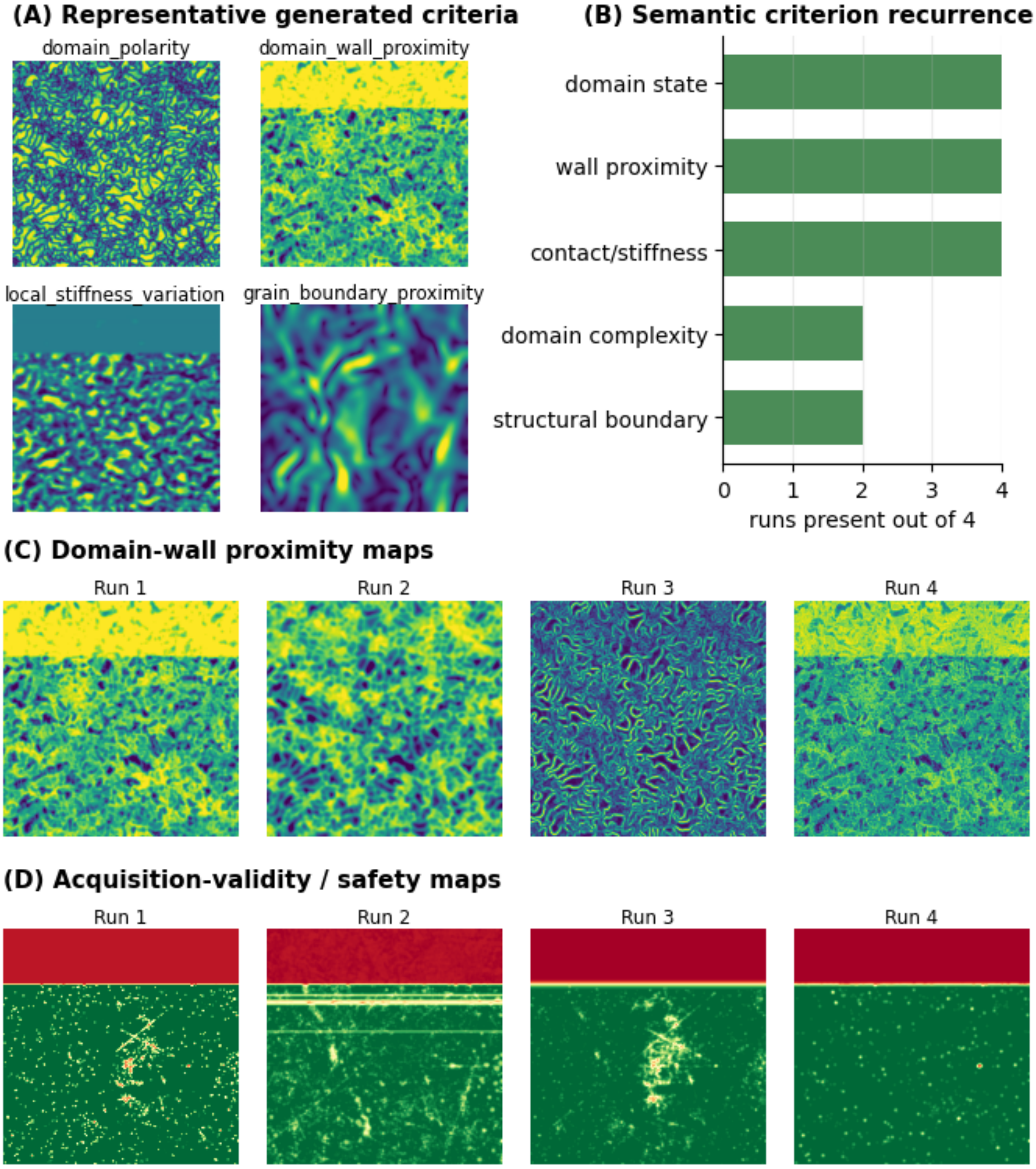

**Figure 4. Agentic generation of task-dependent spatial criteria for PLZT PFM imaging.** (A) Representative spatial criteria generated from the multichannel dataset. (B) Recurrence of semantic criteria across repeated analyses. (C) Repeated generation of domain-wall-proximity maps, showing semantic consistency with variable numerical realization. (D) Acquisition-validity maps identifying regions potentially unsuitable for local spectroscopy.

### 3.2. Deterministic representation and agentic interpretation of local switching spectroscopy

SS-PFM presents a complementary analysis problem. The measured hysteresis response can be reduced to a relatively stable quantitative representation, so numerical descriptor extraction remains deterministic. For each loop, the workflow extracts coercive voltages, loop width and imprint, remanent responses, loop height and area, saturation values, branch noise, quadrature residual, and related descriptors. Repeated processing of the same file therefore returns an invariant numerical input to the subsequent interpretation step.

The agent receives this descriptor vector together with the scientific task and experimental context (Figure 5A). Rather than recalculating the hysteresis parameters, the agent role is to evaluate whether the response provides usable physical evidence. The review considers signal quality, incomplete saturation, phase mixing, asymmetry, offsets, noise, and other non-ideal behavior and returns a structured quality assessment and interpretation.

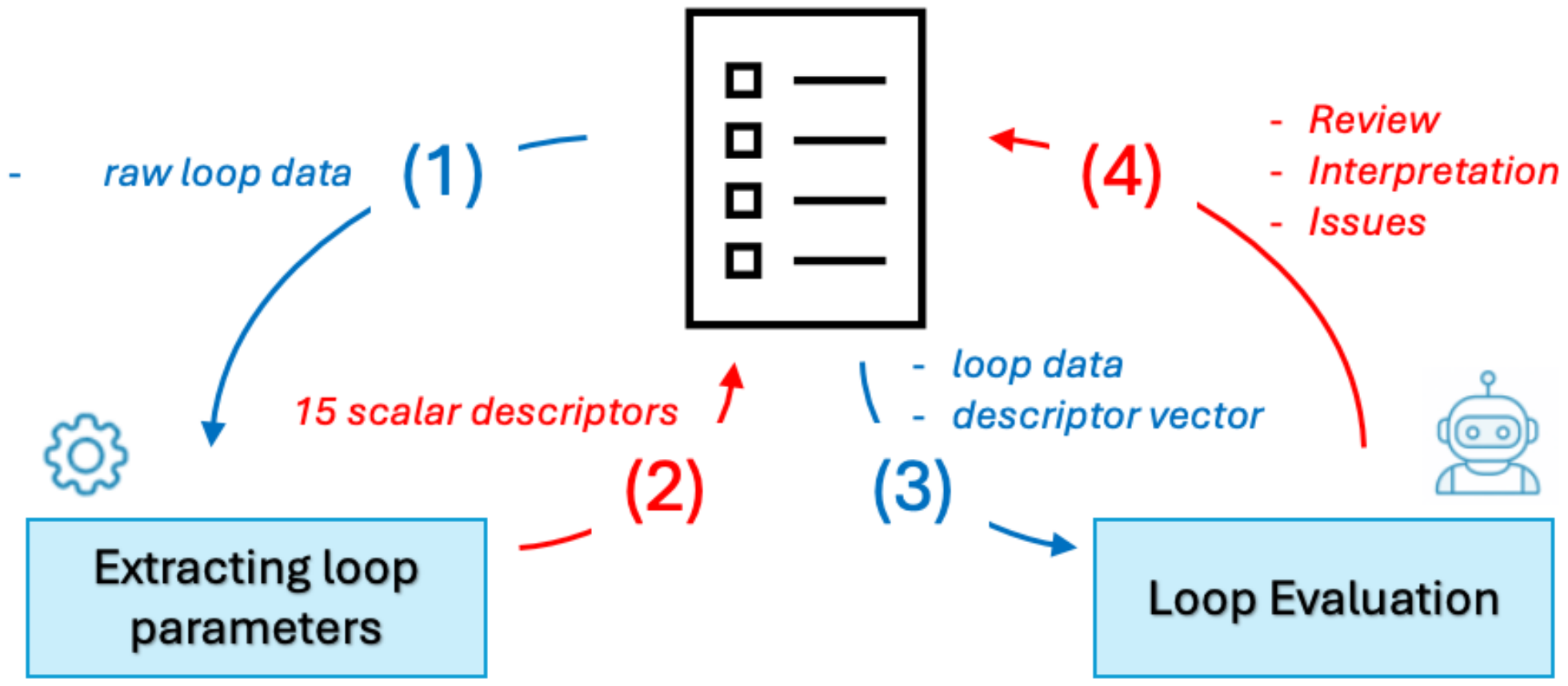

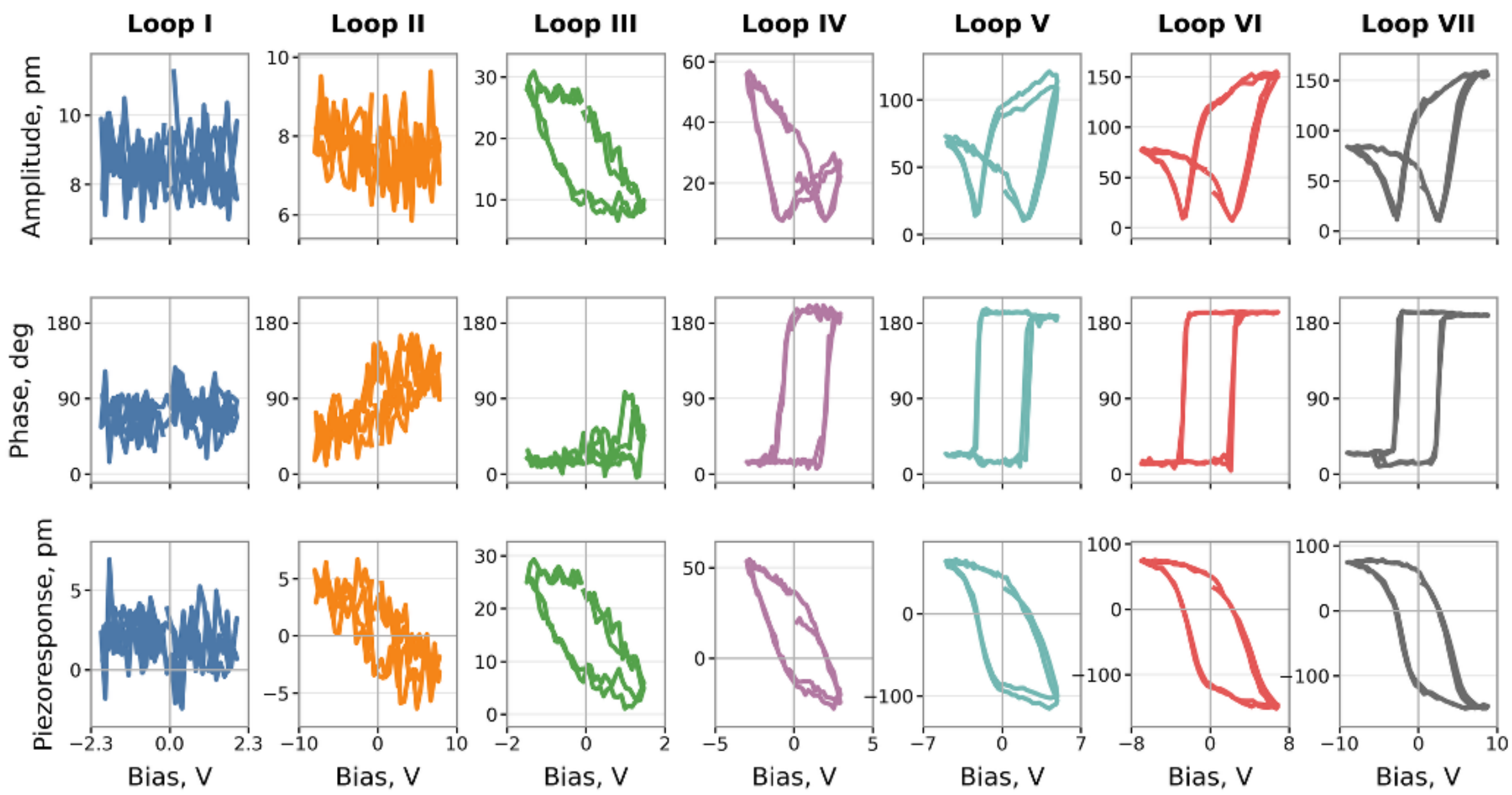

**Figure 5. Deterministic-to-agent workflow and representative off-field hysteresis loops.** (A) Deterministic scalarization converts the measured response to a quantitative descriptor vector before agentic evaluation. (B) Representative off-field DART-PFM amplitude, phase, and piezoresponse loops spanning weak to well-developed switching responses.

Seven loops spanning weak, noisy, intermediate, and well-developed switching responses were used to evaluate this separation (Figure 5B). Each deterministic descriptor vector was evaluated ten times. Agent confidence broadly followed the deterministic quality landscape defined by signal-to-noise ratio and quadrature residual, with better-resolved loops receiving higher confidence and strongly mixed or noisy responses occupying the lower-quality regime (Figure 6A). The deterministic scalarization was invariant by construction, while repeat-to-repeat variation in the agent evaluation remained small compared with differences between loops (Figure 6B). Importantly, confidence did not simply reflect measurement quality, ambiguous intermediate responses received lower confidence than either well-developed loops or measurements that could be rejected more straightforwardly as poor quality.

The repeated evaluations also showed consistent use of the underlying evidence (Figure 6C). Noise and phase-mixing contributions were identified most frequently for weaker responses, whereas asymmetry and other intrinsic loop-shape characteristics remained relevant across a broader range of measurements. Higher-level interpretation categories were similarly stable across repetitions, with well-developed and clearly noise-dominated responses producing the most consistent conclusions and intermediate loops retaining greater interpretive uncertainty (Figure 6D).

The purpose of this analysis is not to establish the agent as a superior numerical loop classifier. Rather, it demonstrates that quantitative measurement processing can remain invariant while contextual interpretation is layered above it as a distinct and independently evaluable operation. A more detailed analysis of the repeatability of the agentic hysteresis-loop evaluation is provided in the Supplementary Information.

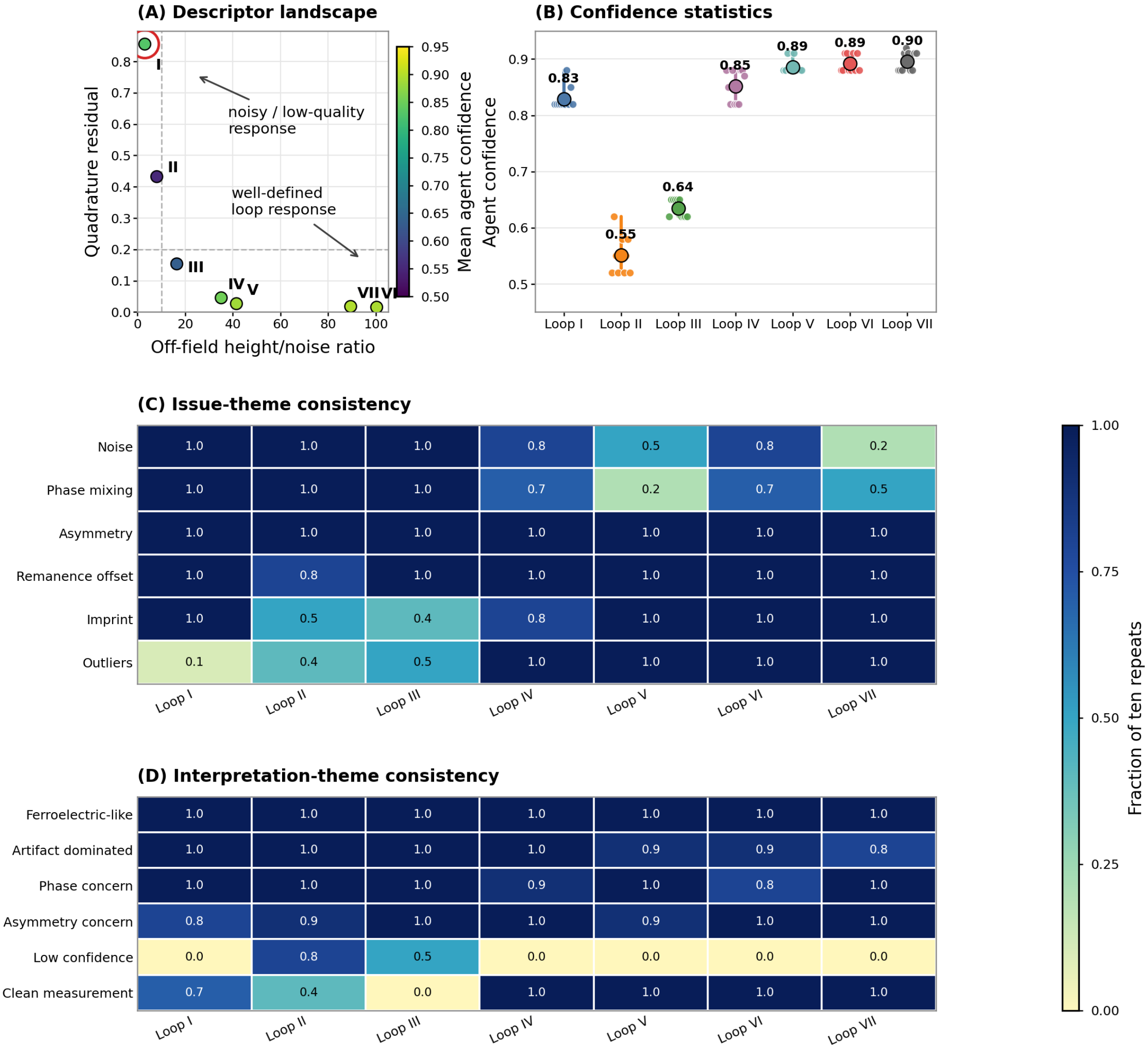

**Figure 6. Agentic evaluation of hysteresis-loop quality and interpretation consistency.** (A) Relation between deterministic quality descriptors and agent confidence. (B) Confidence statistics from repeated evaluations of the same descriptor vector. (C) Consistency of identified evidence factors. (D) Consistency of higher-level interpretation categories.

### 3.4. Adaptive evidence accumulation in the end-to-end autonomous experiment

The complete workflow was evaluated on a PLZT relaxor ferroelectric with a single instruction – *determine how local domain structure affects polarization switching dynamics*. The context stated that this was a DART-PFM measurement on PLZT with complex domain structure. No spatial criteria, measurement locations, spectroscopy waveform, experimental trajectory, or stopping rule were specified for the agent. The system was initialized from the operator's current instrument configuration.

The first 15 × 15 μm scan contained clear polarization-related phase contrast together with topographic scratches and banding in the resonance-frequency channel (Figure 7A). From this scan and the scientific task, the workflow generated three primary spatial coordinates – domain polarization state, domain-wall proximity, and local phase heterogeneity (Figure 7B). An independent acquisition-validity map excluded scratch-like topography and reduced the desirability of rows with anomalous contact-resonance behavior (Figure 7C). After the border margin was applied, 76.2% of the frame remained admissible for spectroscopy.

For each subsequent spectroscopy measurement, the decision node selected a spatial criterion and a sampling strategy, while the final physical coordinate was determined algorithmically from the corresponding criterion and acquisition-validity maps. The resulting trajectory was nonuniform in real space, with several measurements concentrated in regions where the selected criteria reached their most informative values (Figure 7D). This spatial concentration therefore reflects task-conditioned sampling rather than uniform exploration of the image. When projected into criterion space, however, the thirteen measured locations covered low, intermediate, and high values of all three generated descriptors (Figure 7E). The trajectory thus progressively expanded the range of represented local environments even though the corresponding measurement locations were unevenly distributed across the physical field of view.

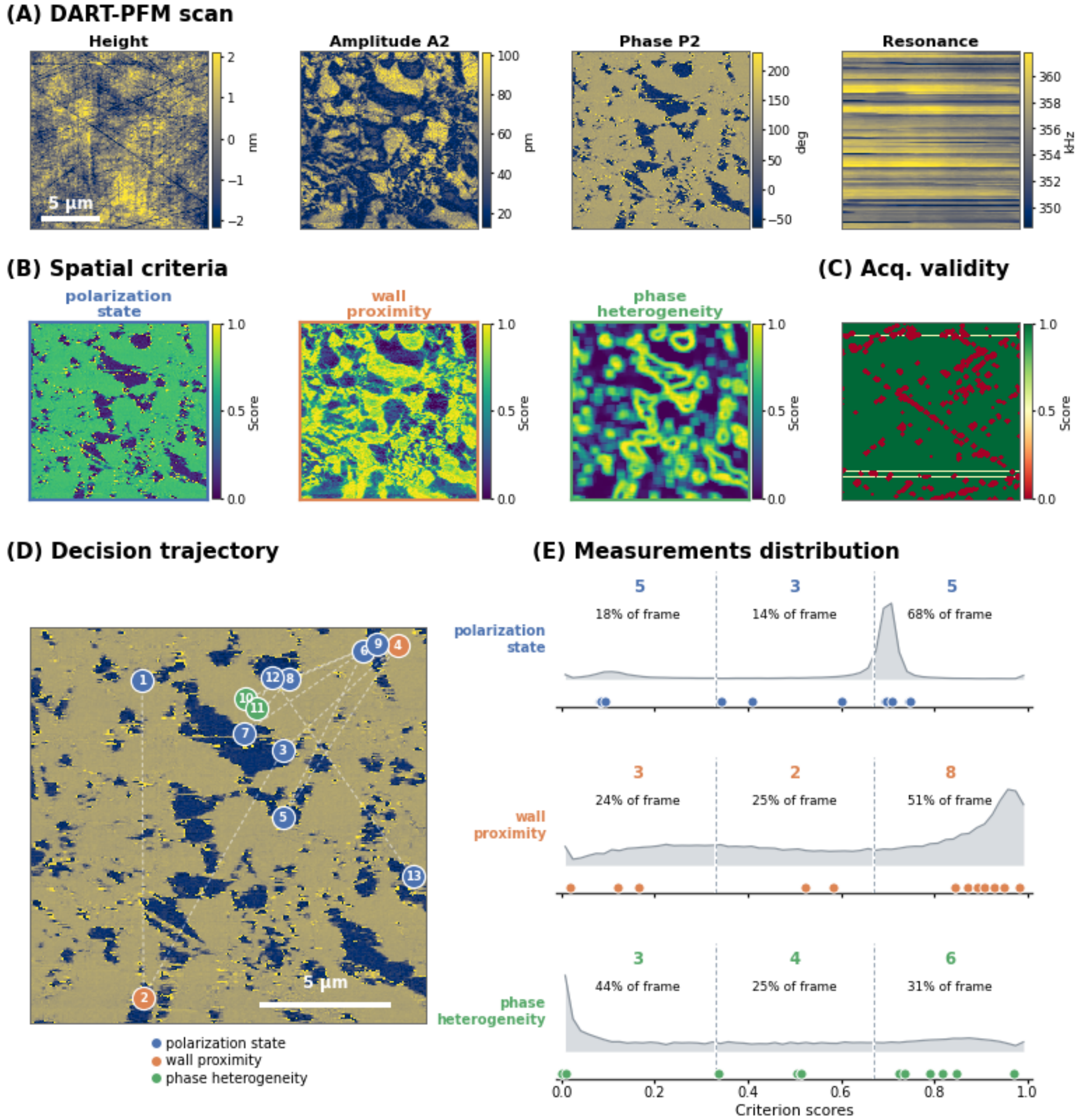

**Figure 7. Autonomous PLZT experiment.** (A) Initial DART-PFM scan channels. (B) Generated task-conditioned spatial-criterion maps. (C) Acquisition-validity map. (D)

Experimental decision trajectory in real space, with loop positions colored by the criterion used for selection. (E) Criterion-space coverage of the measured locations.

The decision node could request another loop, acquire another scan, or terminate. For spectroscopy, it selected both a criterion and a sampling strategy, i.e., max, min, or diverse. It never selected the physical coordinate directly. A deterministic point-selection algorithm converted the requested criterion and strategy into an admissible pixel using the criterion map, the acquisition-validity map, and a penalty against repeatedly sampling nearby sites.

The first measurements bracketed the main axes of the generated criterion space. The system sampled high polarization state, high wall proximity, low polarization state, and low wall proximity, then selected additional measurements intended to fill gaps in the sampled representation (Figure 8A). Spectroscopy parameters were adapted separately (Figure 8B). The first loop used the initial +/-3.5 V bias window. Because this response was noisy and not fully saturated, the next plan increased the sweep to +/-4.0 V, and after incomplete saturation remained, to +/-4.5 V. Once the larger window produced usable switching loops, the waveform was held fixed for the remaining eleven measurements.

This transition is important because the system did not continue optimizing every loop independently. Once a usable waveform had been established, the experimental priority shifted from improving an individual measurement to maintaining comparability among measurements acquired at different microstructural sites. As the trajectory progressed, the novelty of newly sampled locations in criterion space generally decreased, indicating progressive saturation of the accessible representation (Figure 8C). At the same time, the acquired loops remained suitable for interpretation. 11 of 13 off-field loops were classified as good and two as noisy, with agent confidence broadly following the deterministic signal-quality measures (Figure 8D). The measured off-field responses are shown in Figure 8E. As can be seen 13 loops were acquired over 14 decisions, with all three generated criteria sampled across low, intermediate, and high score ranges. The experiment was ultimately terminated by the system itself once the accumulated evidence was judged sufficient.

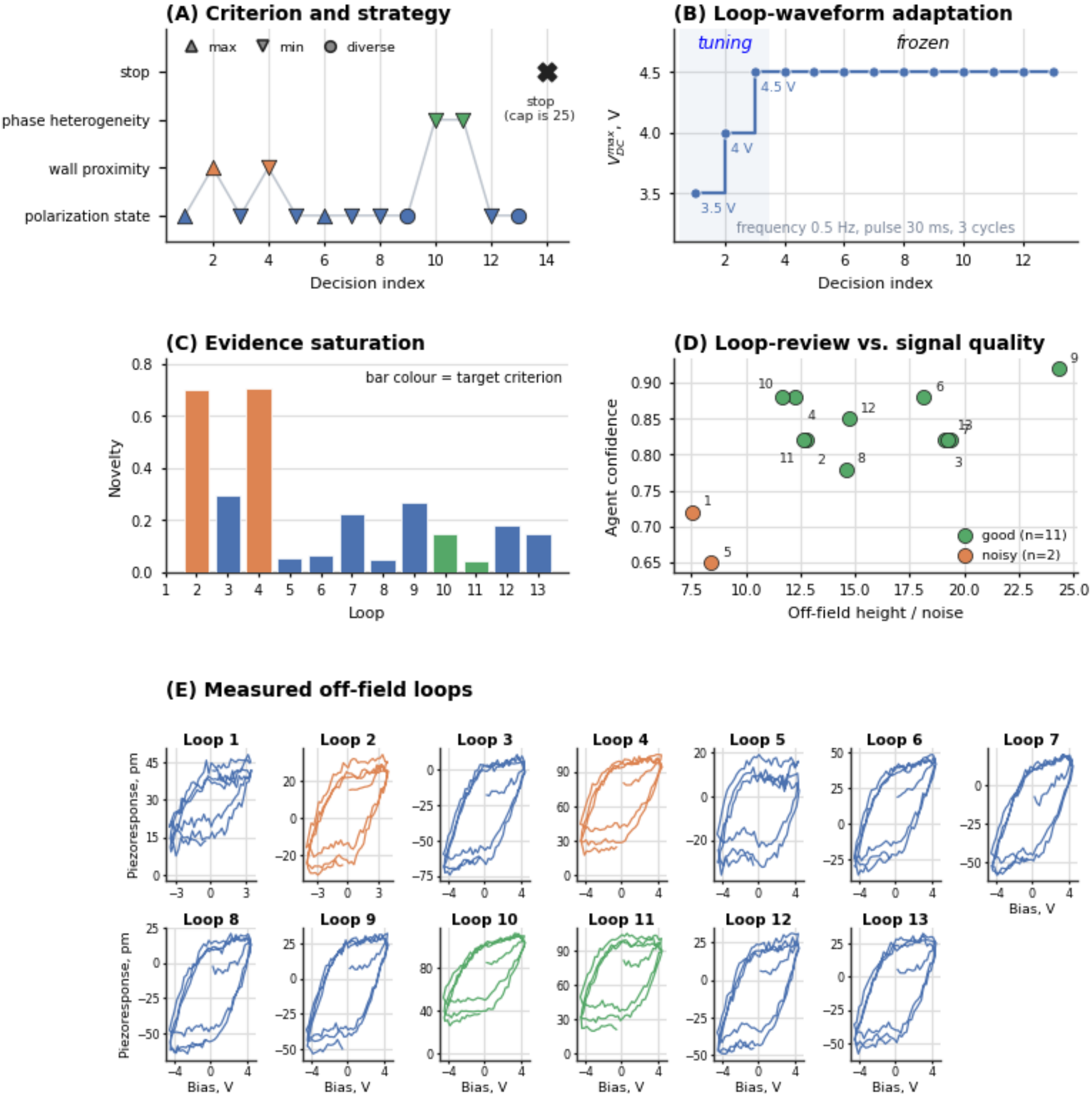

**Figure 8. Agentic decision dynamics and loop evidence.** (A) Criteria and sampling strategies selected during the autonomous trajectory, ending with termination before the hard cap. (B) Loop-waveform adaptation, with early bias tuning followed by fixed conditions for comparison. (C) Declining novelty in criterion space. (D) Agent loop-review confidence versus deterministic signal quality. (E) Measured off-field hysteresis loops, colored by the criterion that selected each site.

### 3.5. Point-selection audit, scientific inference, and experimental identifiability

The archived trajectory allows the deterministic point-selection step and the scientific content of the run to be audited independently of the agent. Reimplementing the coordinate selector from the stored criterion and validity maps reproduced the recorded pixel exactly for 11/13 loop measurements (Figure 9A). The acquisition-validity factor materially affected the trajectory, so removing it while keeping the criterion and spacing terms unchanged altered 10/13 selected locations, and 6 of the relocated sites fell on pixels with validity below 1 (Figure 9B). The validity map therefore acted as an active experimental constraint rather than a cosmetic filter.

The trajectory also revealed an unresolved scientific contrast (Figure 9C). Measurements selected to probe low-polarization-state regions were repeatedly also close to domain walls. The agent recognized that polarization state and wall proximity were therefore not being sampled independently and returned to the low-polarization criterion at decisions 7, 8, and 12 in an attempt to obtain a low-polarization, low-wall-proximity comparison.

Analysis of the complete admissible field of view shows why these attempts could not succeed. No admissible pixel had both polarization-state and wall-proximity scores below 0.33, and the same combination remained absent when both thresholds were increased to 0.40 or 0.50. Among pixels with polarization-state score below 0.33, wall proximity had a minimum of 0.53 and a median of 0.94. The requested comparison therefore did not exist within the available experimental state space (Figure 9C).

The generated descriptors were also coupled more broadly. Wall proximity and local phase heterogeneity correlated at $r = +0.64$ over the admissible frame and at $r = +0.91$ over the 13 measured sites (Figure 9D). Polarization state and wall proximity correlated at $r = -0.46$ over the frame and $r = -0.71$ over the measured locations. Thus, the sampling trajectory can amplify correlations already present in the microstructure, and criterion-space coverage alone does not guarantee that the scientific variables have been independently sampled.

The measured loops nevertheless contained structure-property information, including systematic variations of response offset with polarization state and of loop width across wall-proximity regimes (Figure 9E, F). These trends must be interpreted subject to the coupling between descriptors. The important outcome is therefore not only the set of acquired loops, but an explicit statement of which scientific distinctions the current experiment can and cannot support.

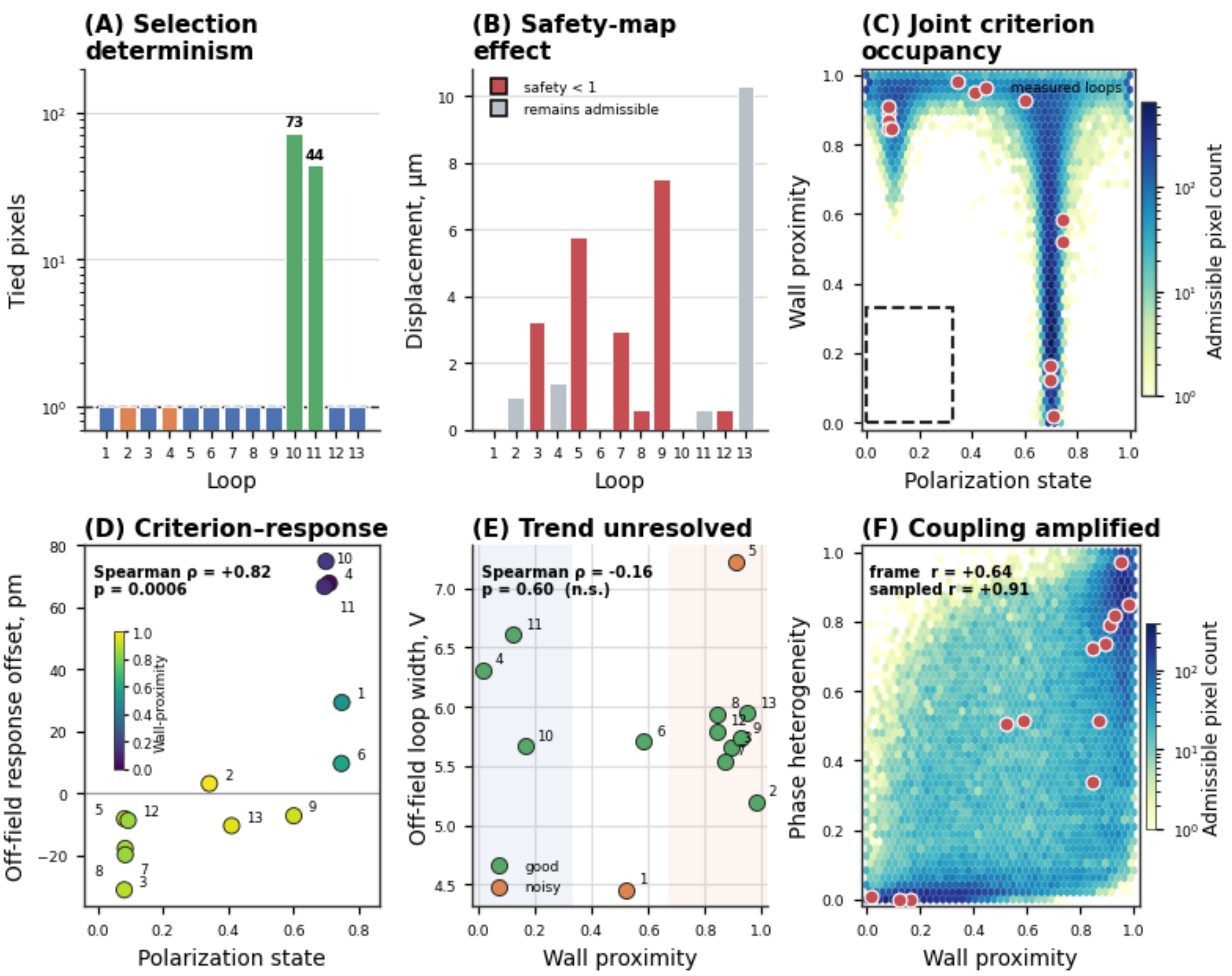

**Figure 9. Audit of point selection and experimental identifiability.** (A) Number of pixels tied at the deterministic maximum for each loop. (B) Change in selected position when the acquisition-validity map is removed; red bars indicate relocations to reduced-validity pixels. (C) Accessible polarization-state/wall-proximity space and measured sites; the dashed region marks the repeatedly targeted low–low combination that is absent from the field of view. (D) Off-field response offset versus polarization state, colored by wall proximity. (E) Off-field loop width versus wall proximity, showing no significant trend. (F) Wall-proximity/phase-heterogeneity occupancy, showing stronger coupling in the sampled sites than across the full field.

The experiment terminated after decision 14, before the hard cap of 25 decisions. The final evidence digest showed that the loops spanned 69% by 90% of the frame, all three criteria had been sampled across their principal score bands, and criterion-space novelty decreased from 0.41 for the first three loops to 0.12 for the last three. The agent also retained the unresolved low-polarization, low-wall-proximity contrast as a limitation. Additional local spectroscopy within the same frame was therefore unlikely either to add substantial criterion-space diversity or to provide the missing independent state.

This distinction is central to scientific autonomy. Experimental convergence does not imply complete scientific resolution. A valid autonomous outcome can instead be recognition that the accumulated evidence supports a bounded conclusion while the remaining ambiguity is not resolvable under the current experimental conditions.

The full decision-making traces for the presented end-to-end experiment are provided in the Supplementary Information and are also publicly available (see Data Availability).

## 4. Perspectives for scalable and adaptive autonomous microscopy

The present implementation uses a general-purpose reasoning model and a predefined experimental graph. A natural extension is to introduce reusable, modality-specific experimental skills derived from previous measurement records.[51] Such records can encode conditional relationships between instrument state, parameter changes, measurement quality, and experimental outcomes, providing grounded priors for future planning without reducing them to a fixed table of recommended settings.

The hierarchical structure also allows a complete PFM workflow to become one subgraph within a larger SPM experiment that can select among PFM, Kelvin probe force microscopy, conductive AFM, mechanical mapping, or other modalities. The same principle can extend to multimodal experiments spanning different instruments. The key requirement is that each subgraph expose a compact structured representation of its principal observations, confidence, limitations, and unresolved questions while retaining references to the underlying data.

A more ambitious extension is to make the workflow structure itself adaptive. In the present system, agents select actions within a validated graph. A higher-level system could instead assemble experimental graphs from libraries of validated measurement, analysis, optimization, safety, and termination primitives. For physical experiments, such generated graphs should still be validated against instrument constraints before execution.

The numerical optimization beneath the agentic layer can likewise become substantially more sophisticated without changing the architecture. The max/min/diverse selector used here was deliberately simple so that the separation between semantic selection and numerical coordinate calculation remained explicit. Once an agent identifies what physical quantity should be explored, Bayesian optimization, active learning, representation learning, expected-information-gain methods, or other acquisition functions can determine where measurements should be performed most efficiently. The resulting hierarchy is not agent versus algorithm – the agent determines what should be learned and why, while specialized algorithms determine how the required evidence is acquired and analyzed efficiently, reproducibly, and within physical constraints.

Realizing this extension will require evaluation beyond whether a multi-agent system completes an experiment. Appropriate benchmarks should measure source attribution, hypothesis diversity, calibration of stated uncertainty, consistency across remated runs, recovery from incorrect immediate interpretations, experimental efficiency, computational cost, and the validity of stopping decisions. Comparisons among single-agent workflows, specialized multi-agent variants and deterministic baselines would establish whether additional deliberative structure provides scientific value rather than additional complexity. Archived experimental trajectories such as those generated here provide a practical substrate for such evaluations because the same evidence can be replayed under controlled ablations without repeatedly operating the physical instrument. The resulting architecture is therefore not agent versus algorithm, nor single agent versus many agents. It is a hierarchy in which literature retrieval, hypothesis construction, criticism, and scientific planning are separated into auditable reasoning roles, while numerical analysis, optimization, validation, and physical execution remain grounded in verifiable procedures. Such a system could progress from autonomous measurement selection toward autonomous scientific inquiry while retaining explicit boundaries among prior knowledge, proposed explanations, acquired evidence, and physically justified conclusions.

## 5. Summary

We developed a hierarchical dual architecture for autonomous physical experimentation in which agentic and algorithmic operations are assigned according to the structure of the scientific task. Agentic reasoning interprets scientific objectives, constructs task-dependent experimental representations, evaluates accumulated evidence, and selects high-level actions. Deterministic algorithms perform quantitative analysis, optimization, coordinate selection, validation, and instrument execution.

In an autonomous PFM experiment, this architecture converted a broad question about domain structure and polarization switching into spatial experimental coordinates, local spectroscopy measurements, adaptive waveform selection, and an evidence-based stopping decision. The experiment also exposed a more consequential form of autonomy, where the system identified that polarization state and wall proximity were confounded in the accessible field of view and that the contrast required to distinguish their effects was absent. Autonomous experimentation therefore need not terminate only when an objective has been optimized; it can also terminate with a bounded scientific conclusion stating what the acquired evidence supports, what remains unresolved, and why the present experiment cannot resolve it. The separation of agentic reasoning, quantitative analysis and deterministic execution provides a reusable framework that the scientific community can adapt to other instruments, modalities, materials and experimental questions.

## 6. Methods

### 6.1 Sample and microscope

Measurements were performed on a 7/65/35 lead lanthanum zirconate titanate (PLZT) relaxor ferroelectric ceramic manufactured by Boston Applied Technologies, Inc. Asylum Research Jupiter atomic force microscope (Oxford Instruments) was used with a probe Multi 75E-G (spring constant ~3 N/m) at ambient conditions. Imaging used dual AC resonance tracking PFM (DART-PFM), in which two drive frequencies straddle the contact resonance so that amplitude and phase remain tracked as the contact stiffness varies.

The microscope was driven through an SPM MCP server running on the instrument PC, so that every command was issued as a validated tool call and the achieved state was read back before the workflow continued. The microscope control is performed through the aespm interface Python library.[52] Tip moves were required to land within one pixel of the requested position, and lateral travel was bounded to ±40 μm from the stage origin. Bias amplitude was bounded to 1-8 V, loop frequency to 0.1-1 Hz, pulse time to 1-100 ms and cycle count to 1-3; scan size was bounded to 0.1-30 μm, line rate to 0.1-1.5 Hz, AC drive to 0.05-5 V, DART bandwidth to 2-30 kHz and DART gain to 50-1500. Any plan violating a bound was rejected and returned for correction before reaching the instrument.

### 6.2 Workflow implementation

The experiment was implemented as a LangGraph state graph over a single shared experiment state holding the task, context, instrument state, measurement records, generated criteria, decisions and remaining budget.[53] Pixel arrays were never held in the state; they were written to disk and passed by reference.

Agentic nodes used Claude Sonnet 4.6 at temperature 0 with structured Pydantic outputs and a 4096-token limit, raised to 8192 for the final summary. These were channel recommendation, criterion construction, loop review, loop planning, the decide node and the campaign summary. Analysis code written by the criterion-construction node was executed in an isolated sandbox under a fixed step cap. All remaining operations were deterministic: file reading through SciFiReaders exposed over MCP, routing by measurement type, loop segmentation and scalarization, map validation, coordinate selection and instrument execution.[54] Transient network failures were retried up to four times with exponential backoff, while validation failures and instrument errors were not retried.

The decide node returned one of three actions – measure a loop, acquire a scan, or stop – together with a frame action (hold, zoom in, zoom out, relocate) for scans, or a named criterion and a sampling strategy (max, min, diverse) for loops. It never returned a coordinate. The deterministic picker converted criterion and strategy into a pixel: the criterion map was averaged over a 3-pixel neighborhood, converted to a desirability map by the strategy, multiplied by the safety map and by a spatial penalty that left 5% desirability at previously measured pixels and recovered over 5 pixels, masked to exclude a 5-pixel border, smoothed with the same neighborhood, and reduced to its arg-max. A hard cap of 25 decisions bounded the run.

### 6.3 Data analysis reproducibility studies

Image-branch reproducibility used four independent runs on the same DART-PFM file (PLZT_ceramics.ibw, six channels, 512 × 512) with identical task and context. Agreement was assessed on the recommended channel and its confidence, on the recurrence of semantic criteria across runs, and on Pearson correlations between flattened criterion, task-score, importance and safety maps, computed both raw and after masking to commonly safe regions.

Loop-branch reproducibility used seven acquired SS-PFM files analysed ten times each, giving 70 runs. The deterministic scalarization is identical by construction, so only the agentic review was repeated; consensus was the modal quality label over ten repeats and agreement the fraction of repeats matching it. Issue and interpretation themes were extracted from the generated text using fixed keyword rules.

### 6.4 Data record

Raw Igor binary wave files were converted to a common representation of channel names, units, array paths, basic statistics and preview images. Numerical arrays were stored as npy and every agentic and deterministic decision as structured json in a per-run folder, so the complete trajectory – scan, criteria, safety map, picks, waveforms, extracted descriptors, reviews and stated reasoning – is recoverable after the fact. All statistics and figures in Sections 3.2-3.5 were regenerated from those folders alone (see Data Availability).

**Acknowledgements**

Development of the agentic control framework and autonomous experimental workflow (BNS, AS, YL, SVK) was supported by the National Science Foundation under Award No. 2523284. Development of the experimental and computational infrastructure enabling the autonomous measurements (SLS, MA) was supported by the NSF PCL ATHENA program under Award No. 2607469. Initial MCP server development (GD) was supported by AI Tennessee program.

**Data Availability**

The agentic system used in this study is available on GitHub: https://github.com/Slautin/SPM_agent_simple (release v.1.0.0). The experimental data, decision-making traces, and the Jupyter notebook required to reproduce the analyses presented in this article are available on Zenodo: https://doi.org/10.5281/zenodo.22136469.

*Supplementary Information*

**Hierarchical automation of scanning probe microscopy through agentic orchestration and algorithmic control**

Boris N. Slautin, Sheryl L. Sanchez, Aidan Swanger, Yu Liu,
Gerd Duscher, Mahshid Ahmadi, Sergei Kalinin

Department of Materials Science and Engineering, University of Tennessee; Knoxville, Tennessee, 37996, USA

## S1. Data record and reproduction

Every run of the workflow writes a self-contained folder. Numerical arrays are stored as **.npy** files and every agentic and deterministic step as structured **.json**, so that the trajectory of an experiment – scan, generated criteria, acquisition-validity map, selected pixels, planned waveforms, extracted descriptors, agent reviews and the stated reasoning behind each decision – can be reconstructed after the fact. Table S1 lists the contents of one folder from the autonomous run of Section 2.3.

The four image-analysis runs of Section S2, the seventy loop-review runs of Section S3 and the autonomous run of Section S4 are stored in this format. All figures and tables below were regenerated from those folders by a single notebook (PLZT_manuscript_figures.ipynb), which also reproduces Figures 3–8 of the main text.

| Path | Content |
|---|---|
| session.json | Run directory, start time, and the full instrument status read at initialization (frame size, pixel count, scan rate, drive voltage, DART frequency and width, gains, stage and probe positions, and the loop-waveform settings in force at the start). |
| channels/scan_NN/ | channel_recommendation.json – the structured channel assessment for each analysis task, with the primary and secondary channels, confidence, reasoning and warnings; preview_grid.png. |
| importance/scan_NN/ | components_0.npy and components_0.json – the generated spatial criteria with their names and rationales; task_map_0.npy – the combined criterion score; safety_0.npy and safety_0.json – the acquisition-validity map and its rationale; importance_map_0.npy – the score after the validity factor; scoring_code_0.py – the code the agent generated to compute the criteria; validation_0.json – its validation record. |
| records/ | One JSON record per acquisition. Image records hold the channel inventory with units, array paths and per-channel statistics. Loop records hold the full deterministic descriptor vector and the agent review of that loop. |
| decisions/decision_NN/ | digest.txt – the experimental state presented to the decide node (see Note S1); decision.json – the chosen action with the stated understanding, open questions and reasoning; pick.json – the selected pixel with the criterion and validity values at that pixel; loop_plan.json – the planned waveform with its diagnosis and rationale. |
| loops/loop_NNN/ | Bias, amplitude, phase and quadrature traces as .npy, for the on-field and off-field branches. |
| summary/ | The final state digest and the closing summary of the run. |

**Table S1.** Contents of one archived run folder. NN indexes the scan or decision; NNN indexes the measured loop.

## S2. Repeatability of the image-analysis branch

### S2.1 Channel recommendation

The channel-recommendation node returned the same primary channel for every one of the six analysis tasks in all four runs (Table S2). The recommendations are consistent with the physical origin of each channel: phase for polarization-related tasks, height for topography-related tasks, and the DART resonance-frequency channel for scanning artifacts. Task-level confidence varied by at most 0.13 across runs, and the overall confidence lay between 0.75 and 0.78 (Figure S2A, B).

| Analysis task | Primary channel | Recurrence | Confidence range over 4 runs |
|---|---|---|---|
| Ferroelectric domain segmentation | Channel_004 (Phase2Retrace) | 4/4 | 0.82–0.85 |
| Ferroelectric domain wall segmentation | Channel_004 (Phase2Retrace) | 4/4 | 0.75–0.78 |
| Grain boundary segmentation | Channel_000 (HeightRetrace) | 4/4 | 0.60–0.60 |
| Crack and scratch detection | Channel_000 (HeightRetrace) | 4/4 | 0.88–0.90 |
| Surface contamination identification | Channel_000 (HeightRetrace) | 4/4 | 0.65–0.78 |
| Scanning artifact identification | Channel_005 (FrequencyRetrace) | 4/4 | 0.85–0.88 |

**Table S2.** Channel recommendation over four repeated image-analysis runs on the same dataset. Recurrence is the number of runs in which the listed channel was returned as primary.

### S2.2 Generated criteria

The names differ between runs, so the criteria were grouped into physical categories before recurrence was counted (Table S3). Three categories – domain state or response, domain-wall proximity, and contact resonance or stiffness – appeared in all four runs. Two further categories, domain complexity and structural boundary, appeared in two runs each. Domain-wall proximity was generated under the same name in all four runs.

| Category | Run 1 | Run 2 | Run 3 | Run 4 |
|---|---|---|---|---|
| Domain state / response | domain polarity | piezoresponse amplitude | piezoresponse amplitude | domain polarity contrast |
| Wall proximity | domain wall proximity | domain wall proximity | domain wall proximity | domain wall proximity |
| Contact resonance / stiffness | local stiffness variation | contact resonance frequency | stiffness heterogeneity | elastic stiffness heterogeneity |
| Domain complexity / disorder | – | local domain complexity | polarization disorder | – |
| Structural boundary | grain boundary proximity | – | – | structural boundary density |

**Table S3.** Criteria generated in each of the four repeated runs, grouped by physical category. Names are those assigned by the agent, with underscores replaced by spaces. Each run produced exactly four criteria.

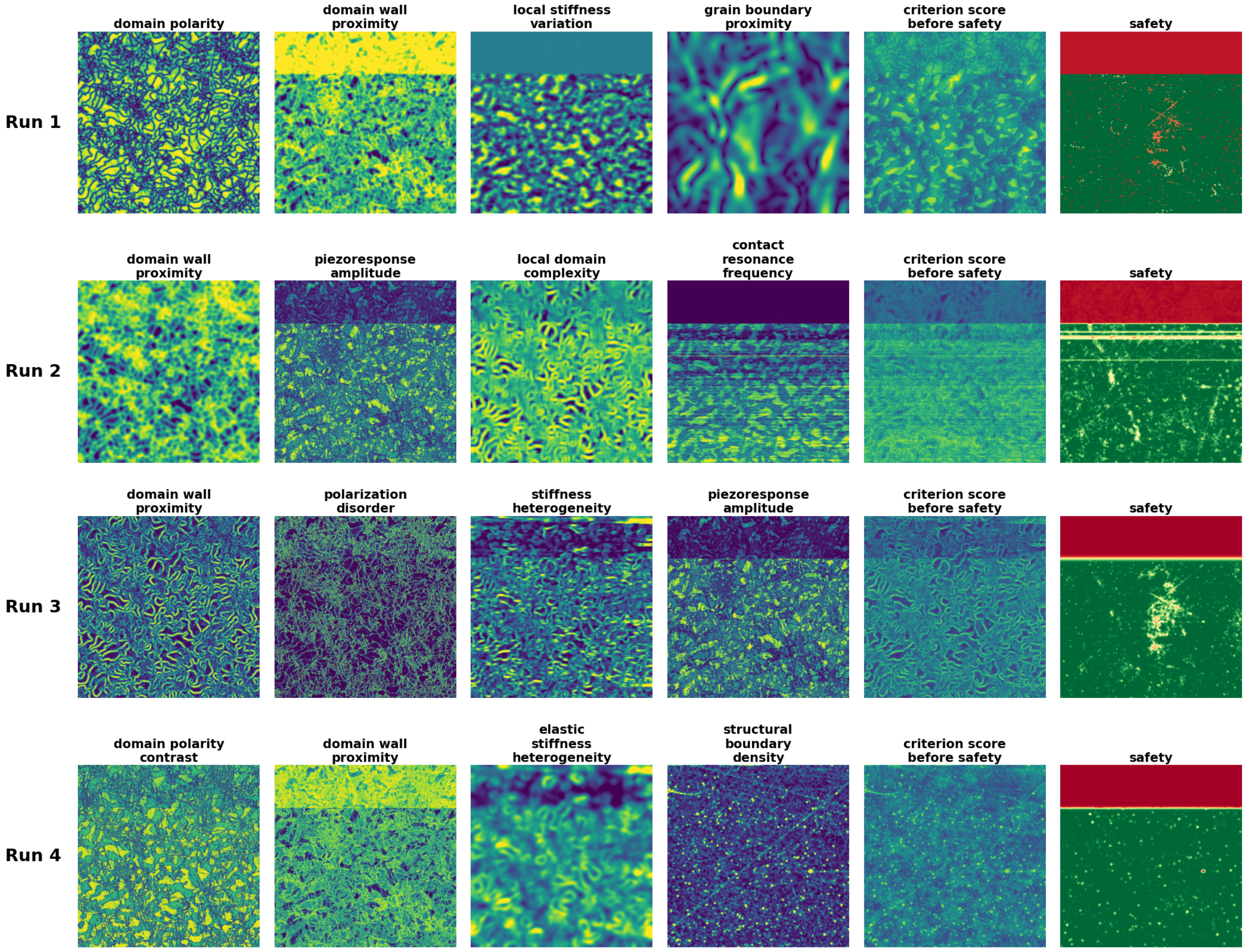

**Figure S1.** All criteria generated in the four repeated image-analysis runs. Rows are runs; the first four columns are the generated criterion maps in the order the agent produced them, followed by the combined criterion score before the acquisition-validity factor is applied and by the acquisition-validity (safety) map. Criterion maps are shown on a common 0–1 scale; safety maps use the red-to-green scale of the main text, with red marking excluded pixels. The horizontal band at the top of the frame is a scan artifact that every run identified independently.

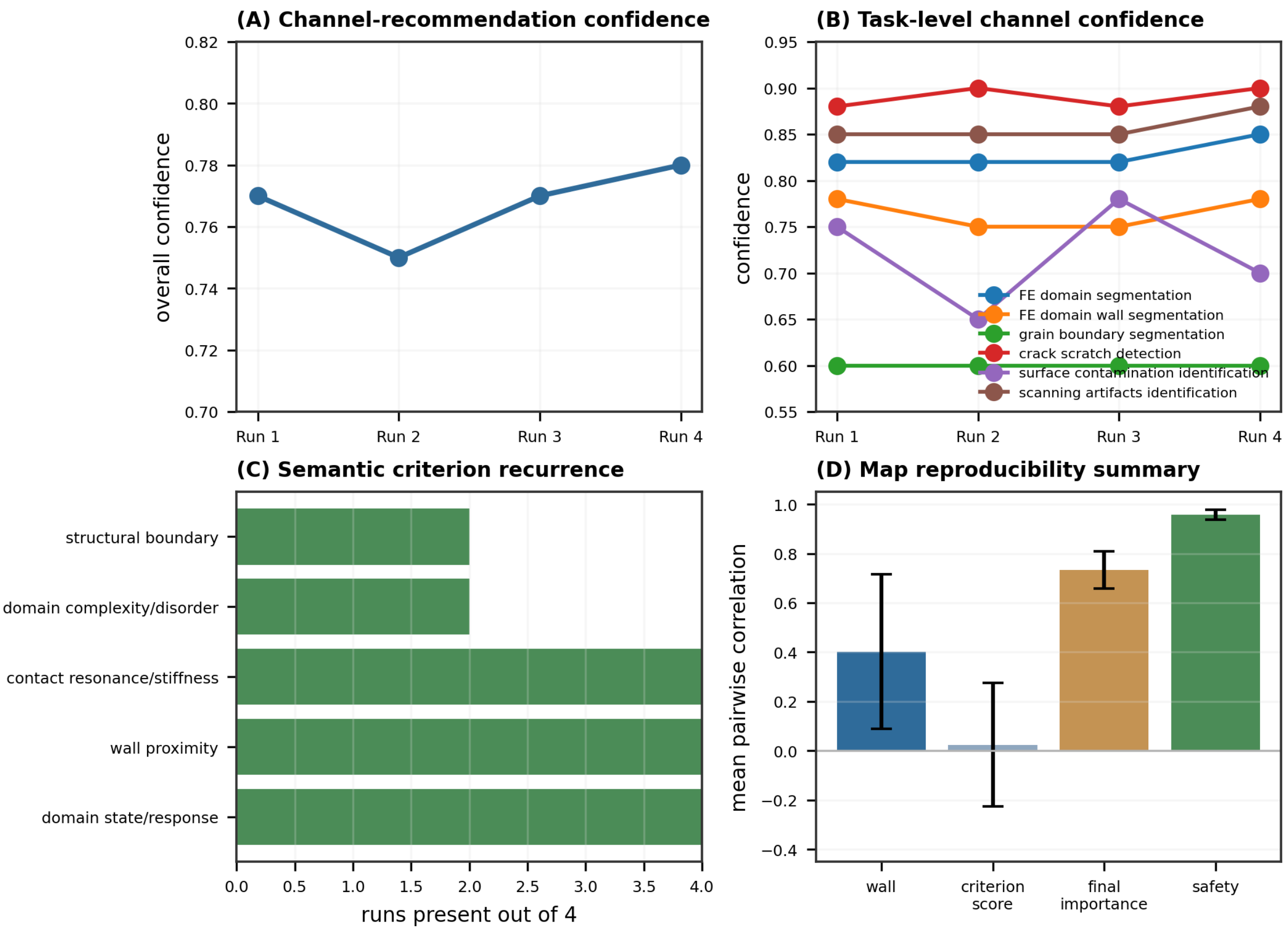

**Figure S2.** Quantitative repeatability of the image branch. (A) Overall channel-recommendation confidence per run. (B) Task-level confidence for the six analysis tasks. (C) Recurrence of the semantic criterion categories over the four runs. (D) Mean pairwise Pearson correlation between the maps of different runs, over the six run pairs; error bars are the standard deviation over those pairs.

### S2.3 Pixel-level agreement between runs

Pixel-level agreement was quantified as the Pearson correlation between flattened maps for each of the six run pairs, computed over the full image and over three restricted regions: pixels that all four runs marked as safe above 0.8, pixels that all four marked as safe above 0.5, and the part of the frame below the artifact band at the top of the image (rows 130 and beyond). The restricted regions are needed because the artifact band is large, is excluded by every run, and therefore contributes strong agreement that is unrelated to the scientific content of the maps.

| Map | Region | Mean r | Range over the six run pairs |
|---|---|---|---|
| Domain-wall criterion | Full image | +0.40 | +0.09 to +0.81 |
| | Common safety > 0.8 | +0.42 | +0.19 to +0.75 |
| | Common safety > 0.5 | +0.42 | +0.19 to +0.76 |
| | Rows ≥ 130 | +0.42 | +0.19 to +0.76 |
| Criterion score before safety | Full image | +0.02 | −0.38 to +0.32 |
| | Common safety > 0.8 | +0.02 | −0.38 to +0.35 |
| | Common safety > 0.5 | +0.02 | −0.37 to +0.33 |
| | Rows ≥ 130 | +0.01 | −0.36 to +0.30 |

| Criterion score after safety | Full image | +0.73 | +0.59 to +0.79 |
| --- | --- | --- | --- |
| | Common safety > 0.8 | +0.01 | −0.37 to +0.36 |
| | Common safety > 0.5 | +0.01 | −0.35 to +0.34 |
| | Rows ≥ 130 | −0.01 | −0.24 to +0.15 |
| Acquisition-validity map | Full image | +0.96 | +0.93 to +0.99 |
| | Common safety > 0.8 | +0.13 | 0.00 to +0.24 |
| | Common safety > 0.5 | +0.18 | 0.00 to +0.47 |
| | Rows ≥ 130 | +0.26 | +0.12 to +0.52 |

**Table S4.** Pixel-level agreement between the four image-analysis runs. Pearson correlation between flattened maps, averaged over the six run pairs, in four regions of the image.

The high full-image agreement of the acquisition-validity maps and of the criterion score after the validity factor is dominated by the region that all runs excluded: once the commonly excluded pixels or the top artifact band are removed, the mean correlation of both maps falls to near zero. The runs therefore agree on which regions to avoid, but not on the pixel-level ranking within the admissible region. Second, the domain-wall criterion retains a modest but non-zero agreement in every region, mean $r \approx 0.42$, indicating that this concept is realized in a partially consistent way even at the pixel level, whereas the composite score is not. This is the quantitative form of the statement in Section 2.2.1 that reproducibility of this branch should be assessed at the level of scientific meaning rather than at the level of individual pixels.

## S3. Repeatability of the loop-analysis branch

Seven measured switching-spectroscopy loops were analysed ten times each, giving seventy runs. The deterministic scalarization and the agentic review are separated in this branch, so the two levels can be tested independently: the scalarization should return identical numbers for every repeat of the same file, and any variability that remains must originate in the interpretation step.

### S3.1 Invariance of the deterministic scalarization

For each loop and each field branch, the range of every numerical descriptor over the ten repeats was computed and the largest range retained. It was exactly zero in all fourteen cases (Table S5), confirming that the agent received an invariant quantitative representation and that the spread reported below is a property of the interpretation alone.

| Loop | Max descriptor range, off-field | Max descriptor range, on-field |
|---|---|---|
| **Loop I** | 0.000 | 0.000 |
| **Loop II** | 0.000 | 0.000 |
| **Loop III** | 0.000 | 0.000 |
| **Loop IV** | 0.000 | 0.000 |
| **Loop V** | 0.000 | 0.000 |
| **Loop VI** | 0.000 | 0.000 |
| **Loop VII** | 0.000 | 0.000 |

**Table S5.** Invariance of the deterministic loop scalarization. For each loop, the largest range of any extracted numerical descriptor over ten repeated analyses of the same file.

### S3.2 Off-field and on-field descriptors

The main text uses the off-field branch, which carries the remanent response after the electrostatic drive is removed. Table S6 and Figure S3 give both branches for the seven loops. In six of the seven loops the off-field branch has both the larger loop width and the higher height-to-noise ratio; the exception is Loop II, which is one of the two weakest measurements. Loop I, the weakest, has an off-field quadrature residual of 0.855, roughly twice its on-field value, consistent with an off-field signal that is dominated by noise rather than by switching.

| Loop | Width off, V | Width on, V | Quad. res. off | Quad. res. on | H/N off | H/N on |
|---|---|---|---|---|---|---|
| Loop I | 0.648 | 0.407 | 0.855 | 0.428 | 3.1 | 2.5 |
| Loop II | 6.471 | 11.539 | 0.432 | 0.466 | 8.1 | 11.0 |
| Loop III | 1.703 | 0.379 | 0.154 | 0.037 | 16.5 | 11.4 |
| Loop IV | 2.848 | 1.100 | 0.046 | 0.022 | 35.1 | 26.8 |
| Loop V | 4.727 | 3.485 | 0.027 | 0.037 | 41.5 | 22.3 |
| Loop VI | 5.252 | 4.494 | 0.016 | 0.034 | 100.4 | 37.3 |
| Loop VII | 5.671 | 5.324 | 0.018 | 0.048 | 89.5 | 36.0 |

**Table S6.** Deterministic descriptors of the seven loops in the off-field and on-field branches. H/N is the height-to-noise ratio.

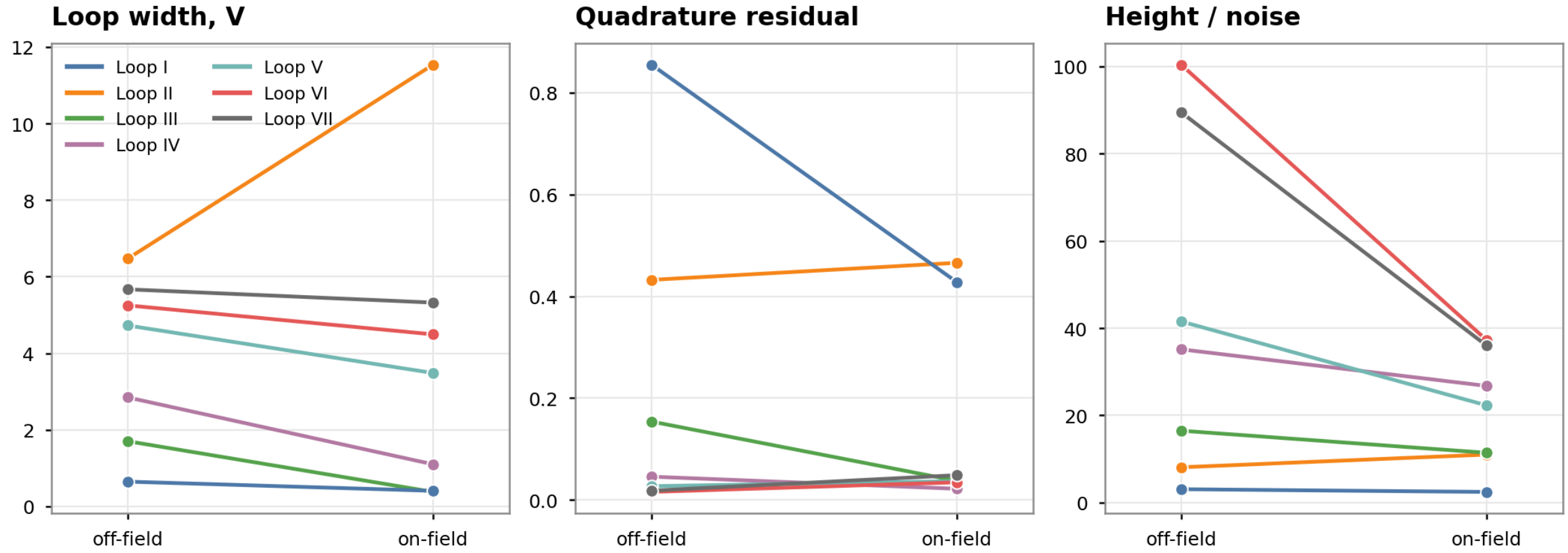

**Figure S3.** Off-field versus on-field descriptors for the seven loops of Figures 5 and 6. Each line connects the two branches of one loop for the loop width, the quadrature residual and the height-to-noise ratio. Colours follow Figure 5.

### S3.3 Repeatability of the agentic review

For each loop, the consensus label is the modal label over the ten repeats and the agreement is the fraction of repeats matching it. The quality label and the ferroelectric-like classification were unanimous for all seven loops (Table S7). The dominant identified issue was unanimous for four loops and reached 0.7–0.9 agreement for the remaining three. Confidence had a standard deviation between 0.012 and 0.032 and a full range of at most 0.10 over ten repeats, so the confidence differences between loops in Figure 6B are large compared with the repeat-to-repeat spread within a loop.

| Loop | Quality | Agree. | Ferroelectric-like | Agree. | Dominant issue | Agree. | Conf. mean ± s.d. |
|---|---|---|---|---|---|---|---|
| Loop I | noisy | 1.0 | no | 1.0 | Noise | 0.8 | 0.829 ± 0.019 |
| Loop II | noisy | 1.0 | yes | 1.0 | Noise | 1.0 | 0.551 ± 0.032 |
| Loop III | noisy | 1.0 | yes | 1.0 | Noise | 1.0 | 0.635 ± 0.015 |
| Loop IV | good | 1.0 | yes | 1.0 | Phase mixing | 0.7 | 0.852 ± 0.027 |
| Loop V | good | 1.0 | yes | 1.0 | Asymmetry | 1.0 | 0.886 ± 0.012 |
| Loop VI | good | 1.0 | yes | 1.0 | Asymmetry | 0.9 | 0.892 ± 0.015 |
| Loop VII | good | 1.0 | yes | 1.0 | Asymmetry | 0.9 | 0.896 ± 0.016 |

**Table S7.** Repeatability of the agentic loop review over ten repeats per loop. Consensus is the modal label; agreement is the fraction of repeats returning it.

Tables S8 and S9 give the full theme matrices behind Figure 6C and 6D. Themes were extracted from the generated text with fixed keyword rules, so the numbers report how often a theme was raised, not how strongly it was argued. Asymmetry and remanence offset were raised in essentially every repeat of every loop, whereas noise and phase mixing were raised in proportion to the measured signal quality. At the interpretation level, the ferroelectric-like reading was raised in every repeat of every loop, and the low-confidence theme appeared only for the two loops with the lowest descriptor quality.

| Theme | I | II | III | IV | V | VI | VII |
|---|---|---|---|---|---|---|---|
| Noise | 1.0 | 1.0 | 1.0 | 0.8 | 0.5 | 0.8 | 0.2 |
| Phase mixing | 1.0 | 1.0 | 1.0 | 0.7 | 0.2 | 0.7 | 0.5 |
| Asymmetry | 1.0 | 1.0 | 1.0 | 1.0 | 1.0 | 1.0 | 1.0 |
| Remanence offset | 1.0 | 0.8 | 1.0 | 1.0 | 1.0 | 1.0 | 1.0 |
| Imprint | 1.0 | 0.5 | 0.4 | 0.8 | 1.0 | 1.0 | 1.0 |
| Outliers | 0.1 | 0.4 | 0.5 | 1.0 | 1.0 | 1.0 | 1.0 |

**Table S8.** Fraction of the ten repeats in which each evidence theme was identified, by loop. Values behind Figure 6C.

| Theme | I | II | III | IV | V | VI | VII |
|---|---|---|---|---|---|---|---|
| Ferroelectric-like | 1.0 | 1.0 | 1.0 | 1.0 | 1.0 | 1.0 | 1.0 |
| Artifact dominated | 1.0 | 1.0 | 1.0 | 1.0 | 0.9 | 0.9 | 0.8 |
| Phase concern | 1.0 | 1.0 | 1.0 | 0.9 | 1.0 | 0.8 | 1.0 |
| Asymmetry concern | 0.8 | 0.9 | 1.0 | 1.0 | 0.9 | 1.0 | 1.0 |
| Low confidence | 0.0 | 0.8 | 0.5 | 0.0 | 0.0 | 0.0 | 0.0 |
| Clean measurement | 0.7 | 0.4 | 0.0 | 1.0 | 1.0 | 1.0 | 1.0 |

**Table S9.** Fraction of the ten repeats in which each interpretation theme was raised, by loop. Values behind Figure 6D.

## S4. The end-to-end autonomous run

This section documents the single autonomous run of Section 2.3: one 15 µm frame at 256 px (58.6 nm/px), thirteen measured loops over fourteen decisions, terminated by the agent at decision 14 against a hard cap of 25. Of the frame, 76.2% was admissible for spectroscopy, defined as validity above 0.5 and outside the border margin; 82.3% of pixels had validity 1.0, 1.9% had 0.5 and 15.8% were excluded.

### S4.1 The state presented to the decide node

The decide node does not see the images. It receives a text digest of the accumulated experimental state, together with the task and context. Note S1 reproduces the coverage part of the digest presented at the final decision. This is the numerical basis on which the stopping decision was taken and is the reason the stated reasoning in Table S10 refers repeatedly to band occupancy and to novelty.

### Note S1. State digest at the final decision (abridged; line breaks adjusted to the page width).

```
FRAME
  area     : 76% of the frame is measurable (safety>0.5, off-border);
             98% of that is still further than 5 px from any loop
  footprint: the loops span 69% x 90% of the frame
  novelty  : first 3 loops 0.41 -> last 3 0.12  (distance in criterion
             space to the nearest earlier loop)
  settled  : the last 3 loops moved the pooled medians by imprint_v 14%,
             loop_width_v 15%, loop_height_m 24% of their IQR
per criterion - 'frame' is the share of pixels in each band, 'measured'
is where the loops landed (low <0.33, high >0.67):
  - domain_polarization_state
      frame   : 17% low / 14% mid / 69% high
      measured: 5 low / 3 mid / 5 high
  - domain_wall_proximity
      frame   : 24% low / 26% mid / 51% high
      measured: 3 low / 2 mid / 8 high
  - local_phase_heterogeneity
      frame   : 45% low / 25% mid / 31% high
      measured: 3 low / 4 mid / 6 high
```

### S4.2 Decision trail

Table S10 lists every decision of the run with the criterion and strategy selected by the agent, the pixel returned by the deterministic point-selection algorithm, the peak bias of the planned waveform, and the criterion values at the selected pixel. The remaining waveform parameters were held at 0.5 Hz, 30 ms pulse time, zero phase and three cycles for all thirteen loops; only the peak bias was adapted, and only over the first three decisions.

| D | Action | Criterion | Strategy | Pixel (y, x) | $V_{DC}$, V | Pol. state | Wall prox. | Phase het. |
|---|---|---|---|---|---|---|---|---|
| 1 | loop | domain_polarization_state | max | (34, 72) | 3.5 | 0.75 | 0.52 | 0.51 |
| 2 | loop | domain_wall_proximity | max | (239, 73) | 4.0 | 0.34 | 0.98 | 0.85 |
| 3 | loop | domain_polarization_state | min | (79, 163) | 4.5 | 0.08 | 0.89 | 0.74 |
| 4 | loop | domain_wall_proximity | min | (11, 237) | 4.5 | 0.71 | 0.02 | 0.01 |
| 5 | loop | domain_polarization_state | min | (123, 163) | 4.5 | 0.08 | 0.91 | 0.79 |
| 6 | loop | domain_polarization_state | max | (15, 215) | 4.5 | 0.75 | 0.58 | 0.51 |
| 7 | loop | domain_polarization_state | min | (68, 138) | 4.5 | 0.09 | 0.87 | 0.51 |
| 8 | loop | domain_polarization_state | min | (33, 167) | 4.5 | 0.08 | 0.85 | 0.72 |
| 9 | loop | domain_polarization_state | diverse | (10, 224) | 4.5 | 0.60 | 0.93 | 0.82 |
| 10 | loop | local_phase_heterogeneity | min | (46, 139) | 4.5 | 0.70 | 0.17 | 0.00 |
| 11 | loop | local_phase_heterogeneity | min | (52, 146) | 4.5 | 0.69 | 0.12 | 0.00 |
| 12 | loop | domain_polarization_state | min | (32, 156) | 4.5 | 0.09 | 0.85 | 0.34 |
| 13 | loop | domain_polarization_state | diverse | (161, 247) | 4.5 | 0.41 | 0.95 | 0.97 |
| 14 | stop | – | – | – | – | – | – | – |

**Table S10.** Decision trail of the autonomous run. The last three columns give the values of the three generated criteria at the selected pixel. The full untruncated reasoning for every decision is archived in decisions/decision_NN/decision.json.

### S4.3 Measured loops

| Loop | Width, V | $V_C$ rise, V | $V_C$ fall, V | Imprint, V | Height, pm | Offset, pm | Quality | Confidence |
|---|---|---|---|---|---|---|---|---|
| 1 | 4.45 | +2.27 | −2.19 | +0.041 | 21.9 | +29.3 | noisy | 0.72 |
| 2 | 5.20 | +2.44 | −2.75 | −0.156 | 42.6 | +3.2 | good | 0.82 |
| 3 | 5.65 | +2.80 | −2.85 | −0.027 | 62.3 | −31.0 | good | 0.82 |
| 4 | 6.31 | +3.11 | −3.20 | −0.046 | 60.8 | +67.8 | good | 0.88 |
| 5 | 7.23 | +3.65 | −3.57 | +0.040 | 38.2 | −8.1 | noisy | 0.65 |
| 6 | 5.71 | +2.71 | −3.00 | −0.148 | 55.1 | +9.7 | good | 0.88 |
| 7 | 5.54 | +2.92 | −2.61 | +0.155 | 56.8 | −17.7 | good | 0.82 |
| 8 | 5.94 | +2.99 | −2.96 | +0.015 | 62.1 | −19.7 | good | 0.78 |
| 9 | 5.73 | +2.86 | −2.87 | −0.001 | 62.8 | −7.2 | good | 0.92 |
| 10 | 5.67 | +2.85 | −2.82 | +0.016 | 55.7 | +74.7 | good | 0.88 |
| 11 | 6.61 | +3.37 | −3.25 | +0.060 | 65.5 | +66.7 | good | 0.82 |
| 12 | 5.79 | +3.03 | −2.77 | +0.129 | 58.0 | −8.6 | good | 0.85 |
| 13 | 5.95 | +3.01 | −2.93 | +0.041 | 67.8 | −10.3 | good | 0.82 |

**Table S11.** Off-field descriptors and agent review of the thirteen loops of the autonomous run. Width is the coercive-voltage separation, $V_C$ rise and fall the two coercive voltages, and offset the vertical displacement of the loop. All thirteen off-field loops were counter-clockwise and all thirteen on-field loops clockwise. Loops 1 and 2 were measured at 3.5 V and 4.0 V respectively; the remaining eleven at 4.5 V.